\documentclass[aps,prd,amsmath,showpacs,nofootinbib]{revtex4}

\def\({\left(}
\def\){\right)}

\newcommand{\beq}{\begin{equation}}
\newcommand{\eeq}{\end{equation}}
\newcommand{\bea}{\begin{eqnarray}}
\newcommand{\eea}{\end{eqnarray}}

\newcommand{\bean}{\begin{eqnarray*}}
\newcommand{\eean}{\end{eqnarray*}}
\newcommand{\bs}{\begin{subequations}}
\newcommand{\es}{\end{subequations}}

\newcommand{\Ref}[1]{(\ref{#1})}

\begin{document}

\title{Scalar wormholes in cosmological setting and their
instability}
\author{Sergey V. Sushkov}%
\email{sergey.sushkov@ksu.ru}%
\affiliation{Department of General Relativity and Gravitation,
Kazan State University, Kremlevskaya str. 18, Kazan 420008,
Russia} %
\affiliation{Department of Mathematics, Tatar State University of
Humanities and Education, Tatarstan str. 2, Kazan 420021, Russia}
\author{Yuan-Zhong Zhang}
\email{zyz@itp.ac.cn}%
\affiliation{Institute of Theoretical Physics, Chinese Academy of
Science, P.O. Box 2735 Beijing 100080, China}


\begin{abstract}
We construct exact nonstatic nonhomogeneous spherically symmetric
solutions in the theory of gravity with a scalar field possessing
the exponential potential. The solution of particular interest
corresponds to the scalar field with negative kinetic energy, i.e.
a ghost, and represents two asymptotically homogeneous spatially
flat universes connected by a throat. We interpret this solution
as a wormhole in cosmological setting. Both the universes and the
wormhole throat are simultaneously expanding with acceleration.
The character of expansion qualitatively depends on the wormhole's
mass $m$. For $m=0$ the expansion goes exponentially, so that the
corresponding spacetime configuration represents two de Sitter
universes joining by the throat. For $m>0$ the expansion has the
power character, so that one has the inflating wormhole connecting
two homogeneous spatially flat universes expanding according to
the power law into the final singularity.

The stability analysis of the non-static wormholes reveals their
instability against linear spherically symmetric perturbations.
\end{abstract}



\maketitle
\section{Introduction}
Wormholes are usually defined as topological handles in spacetime
linking widely separated regions of a single universe, or
``bridges'' joining two different spacetimes
\cite{MorTho,VisserBook}. As is well-known \cite{HocVis}, they can
exist {\em only if} their throats contain an exotic matter which
possesses a negative pressure and violates the null energy
condition. The known classical forms of matter do satisfy the
usual energy conditions, hence wormholes should belong to the
realm of ``unusual'' physics. The search of realistic physical
models providing the wormhole existence represents an important
direction in wormhole physics. Various models of such kind have
been considered in the literature, among them scalar fields
\cite{Ell,Bro,scalarfields}; wormhole solutions in semi-classical
gravity \cite{semiclas}; solutions in Brans-Dicke theory
\cite{Nan-etal}; wormholes on the brane \cite{wormholeonbrane};
wormholes supported by matter with an exotic equation of state,
namely, phantom energy \cite{phantom}, the generalized Chaplygin
gas \cite{chaplygin}, tachyon matter \cite{tachyon}, etc
\cite{footnote,review}.

Recent achievements in observational astrophysics let us to look
at wormholes with the new point of view. The observed acceleration
of the universe means (at least, in the framework of general
relativity) that it is mainly filled by a hypothetical dark
energy: the exotic matter with a positive energy density $\rho>0$
and a negative pressure $p=w\rho$ with $w<-1/3$. The most exotic
form of dark energy is a {\em phantom energy} with $w<-1$
\cite{Cal}, for which the null energy condition is violated. It is
worth to note that values $w<-1$ not only are not excluded but
even are favored by recent observations
\cite{Ton-etal}.

The phantom energy, if exists, can be an appropriate material to
support wormholes. As a simple model for phantom energy ones
consider classical scalar fields with the negative kinetic energy
called {\em ghost} or {\em phantom} scalar fields.\footnote{The
words ``ghost'' and ``phantom'' are often used on equal footing in
papers on gravitation and cosmology. However, it is more correct
to distinguish between ``phantoms'' as ``normal'' fields
possessing an effective equation of state with $w<-1$ and
``ghosts'' as fields with negative kinetic term. (See a discussion
in \cite{BroSta})} Such the fields can play an important role in
cosmology providing the mechanism of the universe acceleration,
also they are able to provide the wormhole existence. Yet in 1973
Ellis \cite{Ell} and, independently, Bronnikov \cite{Bro} found
static spherically symmetric solutions describing wormholes
supported by the scalar field with the negative kinetic energy.
Subsequent investigations \cite{stability,Arm} revealed that such
solutions are stable against small (linear) perturbations. The
stability of static wormholes with ghost fields seems to be
surprising for two reasons, at least. First, different
instabilities arise at boundary surfaces dividing ghost and normal
field behavior which generally transform these surfaces into
singular ones \cite{surface}. Second, serious problems with ghost
fields appear at the quantum level. Actually, the negative kinetic
term leads to the possibility that the energy density may become
arbitrarily negative for high frequency oscillations. From the
quantum field theory point of view this means the possibility of
generating an unlimited amount of particles and antiparticles of
usual positive energy matter fields, accompanied by production of
equal amount of particles and antiparticles of a negative energy
ghost field, i.e. the catastrophic quantum instability of the
vacuum \cite{ghosts}. The above arguments let one suppose that the
stability of static spherically symmetric wormholes supported by
ghost fields is not a general fact.

In this paper we construct and investigate nonstatic spherically
symmetric wormholes in the theory of gravity with a ghost scalar
field. Such the wormhole represents a throat connecting two
universes expanding with an acceleration; the wormhole itself is
evolving together with the expanding universes. The stability
analysis shows that these solutions turn out to be unstable
against linear spherically symmetric perturbations.

The paper is organized as follows. In the section \ref{II} we
write down field equations of the theory of gravity with a scalar
field and briefly consider general properties of static
spherically symmetric wormholes. In the section \ref{III} we
describe the procedure for generating new non-static solutions
being conformally equivalent to ``old'' static ones. New
non-static solutions are analyzed in the section \ref{IV}. It is
shown that the solutions corresponding to the scalar field with
negative kinetic energy represent two asymptotically homogeneous
spatially flat universes connected by a throat, i.e. wormholes in
cosmological setting. In the section \ref{V} we study a stability
of non-static wormholes and show that they turn out to be unstable
against linear spherically symmetric perturbations. The section
\ref{VI} summarizes the results obtained. The appendix contains
some details of derivation of static spherically symmetric
solutions in the theory of gravity with the massless scalar field
minimally coupled to the gravitation field.

\section{Field equations and static spherically symmetric solutions\label{II}}
Consider the theory of gravity with a real scalar field
$\phi$ described by the action
\beq\label{action}
S=\int d^4x\sqrt{-g}\left[R -\epsilon
g^{\mu\nu}\phi_{,\mu}\phi_{,\nu}- 2V(\phi)\right],
\eeq
where $g_{\mu\nu}$ is a metric, $g=\det(g_{\mu\nu})$, $R$ is the
scalar curvature, and $V(\phi)$ is a potential. The value
$\epsilon=+1$ corresponds to an ordinary scalar field with
positive kinetic energy, and $\epsilon=-1$ to a {phantom} field,
i.e., the scalar field with negative kinetic energy.

Varying the action \Ref{action} with respect to $g_{\mu\nu}$ and
$\phi$ yields Einstein equations and the equation of motion of the
scalar field, respectively:
\bs\label{sys}
\bea\label{einstein}
&&R_{\mu\nu}=\epsilon\phi_{,\mu}\phi_{,\nu}+g_{\mu\nu}V(\phi),\\
\label{eqmo} &&\epsilon \nabla^{\alpha}\nabla_{\alpha}\phi=V_\phi,
\eea
\es
where $V_\phi=dV(\phi)/d\phi$.

The static spherically symmetric solution to the Einstein-scalar
equations \Ref{sys} with $V(\phi)\equiv 0$ was first found by
Fisher \cite{Fis} and then was repeatedly rediscovered and
discussed in the literature with various points of view
\cite{Arm,BerLei,Yil,Buc,JanRobWin,Ell,Bro,Wym}. Below we focus
our attention on two general results given by Ellis \cite{Ell} and
Bronnikov \cite{Bro}.

\subsubsection{Ellis solution}

In 1973 Ellis \cite{Ell} exhibited the one-parameter family of
general static spherically solutions divided into three
qualitatively different classes. His result can be reproduced in
the following form:\footnote{For completeness sake we present
details of derivation in the appendix.}

\vskip3pt{\bf Class I.} $\lambda>-1/2$.
\bs\label{statI}
\bea\label{metricI}
&&ds^2=-\(1-\frac{2m}{\delta r}\)^\delta dt^2+\(1-\frac{2m}{\delta
r}\)^{-\delta}dr^2+\(1-\frac{2m}{\delta r}\)^{1-\delta}r^2
d\Omega^2,\\
\label{phiI}
&&\phi(r)=\delta|\lambda|^{1/2}\,\ln\(1-\frac{2m}{\delta r}\),
\eea
\es
where $\delta=(1+2\lambda)^{-1/2}$.

\vskip3pt{\bf Class II.} $\lambda=-1/2$.
\bs\label{statII}
\bea\label{metricII}
&&ds^2=-e^{-2m/r}dt^2+e^{2m/r}[dr^2+r^2d\Omega^2],\\
\label{phiII} &&\phi(r)=-\sqrt{2}\,\frac{m}{r}. \eea
\es

\vskip3pt{\bf Class III.} $\lambda<-1/2$.
\bs\label{statIII}
\bea\label{metricIII}
&&ds^2=-e^{2mu(r)}dt^2+e^{-2mu(r)}[dr^2+(r^2+r_0^2)d\Omega^2],\\
\label{phiIII}&&\phi(r)=\frac{2m|\lambda|^{1/2}}{r_0}
\left(\arctan\frac{r}{r_0}-\frac\pi 2\right) ,
\eea
\es
where  $r_0=m|1+2\lambda|^{1/2}$, and
$u(r)=\frac{1}{r_0}\left(\arctan\frac{r}{r_0}-\frac\pi 2\right).$

Above, $\lambda$ and $m$ are two arbitrary parameters of
integration, and $d\Omega^2=d\theta^2+\sin^2\theta d\varphi^2$ is
the linear element on a unit sphere. It is worth to note that for
all three cases the expression for $\phi$ can be represented in
the unified form:
\beq\label{genphi}
\phi(r)=|\lambda|^{1/2}\ln|g_{tt}|.
\eeq
Note also that $|g_{tt}|=1-2m/r+O(r^{-2})$ in the limit
$r\to\infty$, hence $m$ plays a role of the asymptotical mass for
a distant observer located at $r=\infty$. We will assume that
$m\ge0$.

\subsubsection{Bronnikov solution}
In 1973 Bronnikov \cite{Bro} independently represented another
form for the general static spherically symmetric solution of the
Einstein-scalar equations \Ref{sys} with $V(\phi)\equiv0$. He used
the so-called harmonic radial coordinate $\rho$ such that the
general static spherically symmetric metric
\beq
ds^2=-e^{2\gamma(\rho)}dt^2+e^{2\alpha(\rho)}d\rho^2+e^{2\beta(\rho)}d\Omega^2
\eeq
satisfies the special coordinate condition (the {\em harmonic
gauge})
\beq
\alpha=2\beta+\gamma.
\eeq
In general, the Bronnikov solution reads
\bs
\bea\label{bromet} &&ds^2=-e^{2m\rho} dt^2+\frac{\kappa^2
e^{-2m\rho}}{\sinh^2(\kappa\rho)}\left[\frac{\kappa^2
d\rho^2}{\sinh^2(\kappa\rho)}+
d\Omega^2\right],\\
&&\phi(\rho)=\sqrt{2}\,\big|\kappa^2-m^2\big|^{1/2}\rho.
\label{harmoniccoord}\eea
\es
where $\kappa^2=m^2(1+2\lambda)$. One may easily check that the
relation \Ref{harmoniccoord} for $\phi$ has the form \Ref{genphi}.
In case $\kappa^2=0$ ($\lambda=-1/2$) the metric \Ref{bromet}
reads
\beq
ds^2=-e^{2m\rho}
dt^2+\frac{e^{-2m\rho}}{\rho^2}\left[\frac{d\rho^2}{\rho^2}+
d\Omega^2\right],
\eeq
and in case $\kappa^2<0$ ($\lambda<-1/2$),
\beq
ds^2=-e^{2m\rho} dt^2+\frac{|\kappa|^2
e^{-2m\rho}}{\sin^2(|\kappa|\rho)}\left[\frac{|\kappa|^2
d\rho^2}{\sin^2(|\kappa|\rho)}+ d\Omega^2\right].
\eeq
From the relation \Ref{harmoniccoord} one can see that
$\phi\sim\rho$, i.e. the scalar field, in fact, plays the role of
the harmonic coordinate. Comparing the formula \Ref{harmoniccoord}
with \Ref{phiI}, \Ref{phiII}, and \Ref{phiIII} one may find the
connection between coordinates $\rho$ and $r$.

\section{Generating new non-static solution\label{III}}
Now let us consider the theory \Ref{action} with the scalar
potential $V(\phi)$ in the Liouville, i.e. exponential form
\beq
V(\phi)=V_0 e^{-k\phi}.
\eeq
Note that the exponential potential has been considered in
numerous papers devoted to cosmological models with scalar fields
(see, for instance, \cite{Hal,Bar-etal,LucMat-etal}).
It arises as an effective potential in some supergravity theories
or in Kaluza-Klein theories after dimensional reduction to an
effective four-dimensional theory \cite{Hal}. The exponential
potential also arises in higher-order gravity theories after a
transformation to the Einstein frame \cite{Bar-etal}.

The field equations \Ref{sys} now yield
\bs\label{sys2}
\bea
&&R_{\mu\nu}=\epsilon\phi_{,\mu}\phi_{,\nu}+g_{\mu\nu}V_0 e^{-k\phi},\\
\label{eqmononstat} &&\epsilon
\nabla^{\alpha}\nabla_{\alpha}\phi=-kV_0e^{-k\phi}, \eea
\es%
In 1995 Fonarev \cite{Fon} developed the procedure of generating
new non-static solutions of the system \Ref{sys2}. It was based on
the assumption that a non-static solution is conformally
equivalent to a static vacuum one. As a result Fonarev constructed
the non-static solution being conformally equivalent to the {\em
Class I} static solution \Ref{statI}. Later, in \cite{SusKim} the
new non-static solution being conformally equivalent to the {\em
Class III} static solution \Ref{statIII} was obtained and
analyzed. Now we will present the most general result which can be
expressed as the following\dots

\vskip6pt\noindent {\bf Statement:}\ Let $d\bar s^2$
and $\bar\phi(r)$ be a linear element and a scalar field which
form a static spherically symmetric solution of Einstein-scalar
equations \Ref{sys2} with $V_0=0$, then
\bs\label{state1}
\bea\label{1}
&&ds^2=|\sigma
t|^{4\lambda/(1-2\lambda)}\, d\bar s^2,\\
&&\label{nonstatphi}\phi(t,r)=\bar\phi(r)+\frac{2|\lambda|^{1/2}}{1-2\lambda}\ln|\sigma
t|, \eea
\es
form a non-static solution of \Ref{sys2} with
$V_0={2\sigma^2\lambda(6\lambda-1)}/{(1-2\lambda)^{2}}$ and
$k=|\lambda|^{-1/2}$ provided $\lambda\not=1/2$, and
\bs\label{state2}
\bea\label{2}
&&ds^2=e^{2\sigma t}\,d\bar s^2,\\
&&\phi(t,r)=\bar\phi(r)+\frac{\sigma t}{2},
\eea
\es
form a non-static solution of \Ref{sys2} with $V_0=\sigma^2/2$ and
$k=\sqrt{2}$ provided $\lambda=1/2$, where $\sigma$ is a free
parameter.

\vskip6pt\noindent {\em Proof.} Let $\bar{g}_{\mu\nu}$ and
$\bar{\phi}$ be the `old' static solutions of Einstein-scalar
equations \Ref{sys2} with $V_0=0$. Consider the conformal
transformation of the metric
\beq\label{newg}
g_{\mu\nu}=e^{2\mu(t)}\bar{g}_{\mu\nu},
\eeq
and suppose that at the same time the scalar field transforms as follows
\beq\label{newphi}
\phi=\bar{\phi}+\gamma\mu(t),
\eeq
where $\mu(t)$ is a new indefinite function of $t$, and $\gamma$
is a constant. Using the corresponding transformational properties
of the Ricci tensor:
\bea
R_{00}&=&\bar{R}_{00}-3\ddot{\mu},\nonumber\\
R_{0i}&=&\bar{R}_{0i}+\dot{\mu}\partial_i\ln(g_{00}),\\
R_{ij}&=&\bar{R}_{ij}-(\ddot{\mu}+2\dot{\mu}^2)g_{ij}g^{00},\nonumber
\eea
and taking into account that $\bar{g}_{\mu\nu}$ and $\bar{\phi}$
satisfy the Einstein equations
\beq
\bar{R}_{\mu\nu}=\epsilon\bar{\phi}_{,\mu}\bar{\phi}_{,\nu},
\eeq
it is easy to check that the metric tensor \Ref{newg} and the
scalar field \Ref{newphi} satisfy the field equations \Ref{sys2}
provided the function $\mu(t)$ obeys the following two equations:
\bea
&&\ddot{\mu}+\frac{1-2\lambda}{2\lambda}\,\dot{\mu}^2=0, \\
&&\dot{\mu}^2=-\frac{2\lambda}{1-6\lambda}\,V_0
e^{(2-1/\lambda)\mu},
\eea
and
\beq\label{ak}
\gamma=\epsilon|\lambda|^{-1/2},\quad k=|\lambda|^{-1/2}.
\eeq
In case $\lambda\not=1/2$ these equations are compatible and have
the solution
\beq\label{solmu1}
\mu(t)=\frac{2\lambda}{1-2\lambda}\ln|\sigma(t-t_0)|
\eeq
if and only if
\beq\label{V0-1}
V_0=\frac{2\sigma^2\lambda(6\lambda-1)}{(1-2\lambda)^{2}},
\eeq
where
$\sigma$ and $t_0$ are free parameters. Analogously, in case
$\lambda=1/2$ the solution reads
\beq\label{solmu2}
\mu(t)=2\sigma(t-t_0),
\eeq
and
\beq\label{V0-2}
V_0=\sigma^2/2.
\eeq
Now taking into account Eqs. \Ref{newg}, \Ref{newphi} and making
the rescaling $t\to t+t_0$ we arrive at \Ref{state1} if
$\lambda\not=1/2$, and at \Ref{state2} if $\lambda=1/2$.

To complete the proof we consider the scalar field equation
\Ref{eqmo}. Substituting the expression \Ref{newphi} into
\Ref{eqmo} and taking into account that $\bar{\phi}$ satisfies the
equation $\bar{\nabla}^{\alpha}\bar{\nabla}_{\alpha}\bar{\phi}=0$
we find
\beq\label{eqmomu}
{\epsilon\gamma A^{-1}
e^{-2\mu}}\,(\ddot{\mu}+2\dot{\mu}^2)=kV_0e^{-k(\bar\phi+\gamma\mu)}.
\eeq
As is easy to check straightforwardly, this equation is valid for
$\bar\phi=|\lambda|^{1/2}\ln|g_{tt}|$ and $\mu(t)$, $\gamma$, $k$,
and $V_0$ given by the relations
(\ref{ak}-\ref{V0-2}).\footnote{Notice also that the equation
\Ref{eqmomu} coincides with the $tt$-component of
Einstein's equations.}

\section{Wormholes in cosmological setting\label{IV}}
Now let us analyze the non-static solutions (\ref{1}) and
(\ref{2}) found in the preceding section. First we represent them
in the following unified form:
\beq\label{gennonstatmet}
ds^2=a^2(t)\left[-Adt^2+A^{-1}{dr^2}+Br^2 d\Omega^2\right],
\eeq
where
\beq
a(t)=\left\{
\begin{array}{ccl}
|\sigma t|^{2\lambda/(1-2\lambda)},&\mbox{if}&\lambda\not=1/2,\\
e^{\sigma t},&\mbox{if}&\lambda=1/2,
\end{array}
\right.
\eeq
and functions $A(r)$ and $B(r)$ are defined by the corresponding
static solutions \Ref{metricI}, \Ref{metricII}, or
\Ref{metricIII}, representing the Class I, II, or III,
respectively. Note that in the limit $r\to\infty$ both $A\to1$ and
$B\to1$ for all Classes, and so the metric \Ref{gennonstatmet}
describes in the asymptotic $r=\infty$ an homogeneous spatially
flat universe:
\beq
ds^2=a^2(t)\left[-dt^2+{dr^2}+r^2 d\Omega^2\right].
\eeq
Using the proper time $T=\pm \int a(t)dt$ gives
\beq\label{nonstatmet}
ds^2=-dT^2+b^2(T)\left[{dr^2}+r^2d\Omega^2\right],
\eeq
with
\beq
b(T)=\left\{
\begin{array}{ccl}
\sigma T,&\mbox{if}&\lambda=1/2,\\
\displaystyle\left|\frac{\sigma
T}{1-2\lambda}\right|^{2\lambda},&\mbox{if}&\lambda\not=1/2,\
|\lambda|<\infty,\\
e^{\sigma T},&\mbox{if}&|\lambda|=\infty.
\end{array}
\right.
\eeq
The metric \Ref{nonstatmet} explicitly describes an expanding
universe with the scale factor $b(T)$. Defining the acceleration
parameter $\beta=\ddot{b}b/\dot{b}^2$ we find
\beq
\beta=\frac{2\lambda-1}{2\lambda}.
\eeq
The parameter $\beta$ is negative for $\lambda\in(0,\frac12)$, and
so, in this case, the universe is expanding with deceleration, and
$\beta$ is positive for
$\lambda\in(-\infty,0)\cup(\frac12,\infty)$, that is, the universe
is expanding with acceleration.

Further let us focus on the case of particular interest, i.e. {\em
nonstatic wormholes}. In this case $\lambda<-1/2$, and the
solution is given by formulas \Ref{state1} and \Ref{statIII}.
Taking into account the relation $r_0=m|1+2\lambda|^{1/2}$, we can
represent the nonstatic wormhole metric as follows
\beq\label{nonstatwh}
ds^2=|\sigma
t|^{-\frac{2(m^2+r_0^2)}{2m^2+r_0^2}}\left\{-e^{2mu(r)}dt^2+e^{-2mu(r)}[dr^2+(r^2+r_0^2)d\Omega^2]\right\},
\eeq
where $t\in(-\infty,0)$, $r\in(-\infty,\infty)$, and
$u(r)=\frac{1}{r_0}\left(\arctan\frac{r}{r_0}-\frac\pi 2\right).$
The corresponding solution for the scalar field now reads
\beq\label{nonstatwhphi}
\phi(t,r)=\sqrt{2(m^2+r_0^2)}\left[u(r)+\frac{m\ln|\sigma
t|}{2m^2+r_0^2}\right],
\eeq
and the potential $V(\phi)=V_0 e^{-k\phi}\equiv W(t,r)$ takes the
following form:
\beq\label{W}
W(t,r)=\frac{\sigma^2(m^2+r_0^2)(4m^2+3r_0^2)}{(2m^2+r_0^2)^2}
\exp\left\{-2m\left(u(r)+\frac{m\ln|\sigma
t|}{2m^2+r_0^2}\right)\right\}.
\eeq
The solution \Ref{nonstatwh}, \Ref{nonstatwhphi} depends on three
parameters $m$, $r_0$, and $\sigma$. Depending on a value of $m$
it will be convenient to consider separately different cases:

\vskip6pt\noindent{\bf A.} $m=0$. In this case the solution
\Ref{nonstatwh}, \Ref{nonstatwhphi} takes the especially simple
form:
\beq\label{metm0}
ds^2=|\sigma t|^{-2}\{-dt^2+dr^2+(r^2+r_0^2)\,d\Omega^2\},
\eeq
\beq
\phi(r)=\sqrt{2}r_0u(r).
\eeq
Note that in this case the scalar field $\phi$ does not depend on
the time coordinate $t$, though the metric \Ref{metm0} is
non-static. The potential \Ref{W} becomes to be constant:
\beq
W(r,T)\equiv{3\sigma^2},
\eeq
and corresponds, in fact, to the positive cosmological constant
$\Lambda=3\sigma^2$ in the action \Ref{action}. Using the proper
time $T$:
\beq
-\sigma T=\ln|\sigma t|,
\eeq
yields
\beq\label{metricm0}
ds^2=-dT^2+e^{2\sigma T}[dr^2+(r^2+r_0^2)\,d\Omega^2],
\eeq
It is easy to see that at each moment of time the metric
\Ref{metricm0} coincides asymptotically (i.e. in the limit
$r\to\pm\infty$) with the de Sitter one, and an intermediate
region represents a throat connecting these asymptotically de
Sitter regions. Thus, the spacetime \Ref{metricm0} is a wormhole
joining two de Sitter universes. The instant radius of the throat
is equal to the minimal radius of two-dimensional sphere,
$r_{th}=e^{aT}r_0$; we see that it grows exponentially with time.
Let us calculate now the scalar curvature:
\beq
R=12\sigma^2-\frac{2r_0^2e^{-2\sigma T}}{(r^2+r_0^2)^2}.
\eeq
In the limit $r\to\pm\infty$ as well as in the limit $T\to\infty$
the scalar curvature has the De-Sitter value $R_{DS}=12\sigma^2$,
while at $T=-\infty$ the scalar curvature is singular. This
singularity has a clear geometrical interpretation. Namely, at
each moment of time the throat is represented as the 2D sphere of
minimal radius. In the limit $T\to-\infty$ the radius of sphere
$r_{th}=e^{\sigma T}r_0$ tends to zero, the curvature of sphere
goes to infinity, and the corresponding spacetime scalar curvature
$R$ becomes to be singular.

It is worth also noting that a metric of the kind of
\Ref{metricm0} was first introduced {\em a priori}\ by Roman
\cite{Rom:93}, who explored the possibility that inflation might
provide a mechanism for the enlargement of submicroscopic, i.e.,
Planck scale wormholes to macroscopic size.

\vskip6pt\noindent{\bf B.} $m>0$. In this case the solution is
described by the general formulas \Ref{nonstatwh},
\Ref{nonstatwhphi}. Introducing the proper time coordinate by the
relation:
\beq
|\sigma t|^{-\frac{m^2+r_0^2}{2m^2+r_0^2}}=\left|\tilde\sigma T
\right|^{-\zeta},
\eeq
where $\tilde\sigma=\left(2+\frac{r_0^2}{m^2}\right)\sigma$ and
$\zeta=1+\frac{r_0^2}{m^2}$, we can rewrite the metric
\Ref{nonstatwh} in the following form
\beq\label{powerlaw}
ds^2=-e^{2mu(r)}dT^2+\left|\tilde\sigma T
\right|^{-2\zeta}e^{-2mu(r)}[dr^2+(r^2+r_0^2)d\Omega^2],
\eeq
In two asymptotical regions $r\to\pm\infty$ the last metric
describes homogeneous spatially flat universes:\footnote{In order
to obtain Eq. \Ref{asymet} in the region $r\to-\infty$ one should
take into account that $e^{2mu(r)}\to 1$ at $r\to\infty$, and
$e^{2mu(r)}\to e^{2\pi m/r_0}$ at $r\to-\infty$, and make an
appropriate rescaling.}
\beq\label{asymet}
ds^2=-dT^2+|\tilde\sigma T|^{-2\zeta}\left[d{r}^2+{r}^2
d\Omega^2\right],
\eeq
with the scale factor $a(T)=|\tilde\sigma T|^{-\zeta}$ and the
scalar curvature
$$R =\frac{6\zeta(2\zeta+1)}{T^2}.$$
The corresponding Hubble parameter $H=\dot{a}/a$ is equal to
$\zeta|T|^{-1}$, and the acceleration parameter
$\beta=\ddot{a}a/\dot{a}^2$ is
$$\beta=\frac{\zeta+1}{\zeta}=\frac{2m^2+r_0^2}{m^2+r_0^2},$$
hence the universes are expanding with an acceleration into a
``final'' singularity at $T=0_-$. The intermediate region
$-\infty<r<\infty$ represents a wormhole connecting two universes.
The instant radius of the wormhole's throat is equal to the
minimal radius of two-dimensional sphere that is achieved at $r=m$
and equal to $ r_{th}=|\tilde\sigma T
|^{-2\zeta}e^{-mu(m)}(m^2+r_0^2)^{1/2}$. It is seen that the
throat radius grows according to the power law in the course of
time.

\section{Stability analysis\label{V}}
The stability of static wormholes supported by phantom scalar
fields was investigated in the literature \cite{stability,Arm}. It
was shown that such the wormholes turn out to be stable against
linear spherically symmetric perturbations.

In this section we will study small (linear) spherically symmetric
perturbations of the non-static wormhole solution obtained above.
For this aim we consider the field perturbation
$\phi\to\phi+\delta\phi$ and the perturbed metric
\beq\label{perturbedmetric}
ds^2=|\sigma
t|^{-\frac{2(m^2+r_0^2)}{2m^2+r_0^2}}\left[-e^{2u(r)}(1+\delta\gamma)dt^2+e^{-2u(r)}(1+\delta\alpha){dr^2}+
e^{-2u(r)}(1+\delta\beta)(r^2+r_0^2)d\Omega^2\right],
\eeq
where the perturbations $\delta\phi$, $\delta\alpha$,
$\delta\beta$, and $\delta\gamma$ are functions of $t$ and $r$. In
perturbation analysis there is the so-called gauge freedom, i.e.
that of choosing the frame of reference and the coordinates of the
perturbed space-time. Let us choose the following gauge:
\beq
\frac{2m^2+r_0^2}{m^2+r_0^2}\,e^{4u(r)}\partial_r(-\delta\alpha+2\delta\beta+\delta\gamma)
=\frac{m}{t}\,\partial_t(\delta\alpha+2\delta\beta-\delta\gamma)
+\frac{2m(4m^2+3r_0^2)}{t^2(2m^2+r_0^2)}\,\delta\gamma.
\eeq
In this and only in this case the scalar equation \Ref{eqmo} for
$\delta \phi$ decouples from the other perturbation equations and
reads
\beq\label{dphi}
e^{4u(r)}\left[\partial^2_r\delta\phi+\frac{2r}{r^2+r_0^2}\,\partial_r\delta\phi\right]=
\partial^2_t\delta\phi-\frac{2(m^2+r_0^2)}{t(2m^2+r_0)}\,\partial_t\delta\phi
-\frac{2m^2(4m^2+r_0^2)}{t^2(2m^2+r_0^2)^2}\, \delta\phi.
\eeq
Separating the variables in Eq. \Ref{dphi}:
$\delta\phi=\Theta_\omega(t)\Phi_\omega(r)$, yields
\beq\label{spatialperturb}
\Phi_\omega''+\frac{2r\Phi_\omega'}{r^2+r_0^2}+\omega^2
e^{-4u}\Phi_\omega=0,
\eeq
and
\beq\label{temporalperturb}
\ddot\Theta_\omega-\frac{2(m^2+r_0^2)}{t(2m^2+r_0)}\,\dot\Theta_\omega
+\left[\omega^2-\frac{2m^2(4m^2+r_0^2)}{t^2(2m^2+r_0^2)^2}\right]\Theta_\omega=0,
\eeq
where $\omega^2$ is the constant of separation. The equation
\Ref{spatialperturb} describes a spatial distribution of
perturbations. Its asymptotical solution is
\beq\label{sas1}
\Phi_\omega(r)|_{r\to\infty}= C_1\frac{\sin(\omega
r)}{r}+C_2\frac{\cos(\omega r)}{r},
\eeq
and
\beq\label{sas2}
\Phi_\omega(r)|_{r\to-\infty}= \tilde C_1\frac{\sin(\tilde\omega
r)}{r}+\tilde C_2\frac{\cos(\tilde\omega r)}{r},
\eeq
where $\tilde\omega=e^{2\pi m/r_0}\omega$. In case $\omega^2<0$
the perturbations \Ref{sas1} and \Ref{sas2} diverge at
$|r|=\infty$, and so this case is unphysical. Therefore, we will
analyze the equation \Ref{temporalperturb}, which describes an
evolution of perturbations, assuming that $\omega^2\ge0$. A
general solution of Eq. \Ref{temporalperturb} reads
\beq
\Theta_\omega(t)=|t|^{\frac{4m^2+3r_0^2}{2(2m^2+r_0^2)}}\left[D_1
Y_\nu(\omega |t|)+D_2 J_\nu(\omega |t|)\right],
\eeq
where $J_\nu$, $Y_\nu$ are Bessel functions, $D_1$, $D_2$ are
constants of integrations, and
\beq
\nu=\left[\frac{3(4m^2+r_0^2)(4m^2+3r_0^2)}{4(2m^2+r_0^2)^2}\right]^{1/2}.
\eeq
Consider the behavior of $\Theta_\omega(t)$ in the limit
$t\to0$.\footnote{Let us remind ourselves that $t\in(-\infty,0)$,
so that the ``arrow of time'' is directed from $-\infty$ to $0$,
and $t=0$ corresponds to the distant future.} Taking into account
that near zero $J_\nu(z)\sim |z|^\nu$ and $Y_\nu(z)\sim
|z|^{-\nu}$ (see \cite{AbrSte}) we find
\beq
\Theta_\omega(t)|_{t\to0}\sim D_1|\omega t|^{\nu_-} +D_2|\omega
t|^{\nu_+},
\eeq
where
\beq
\nu_\pm=\frac{4m^2+3r_0^2}{2(2m^2+r_0^2)}\pm\nu=
\frac{4m^2+3r_0^2}{2(2m^2+r_0^2)}\left[1\pm\sqrt{\frac{3(4m^2+r_0^2)}{4m^2+3r_0^2}}\right].
\eeq
It is seen that $\nu_-$ is negative, hence $\Theta_\omega(t)$
behaves near zero as $|\omega t|^{\nu_-}$. Therefore the ratio
$\delta\phi/\phi$, where the non-perturbed solution $\phi$ is
given by \Ref{nonstatwhphi}, is diverging at $t=0$. Physically
this means that linear scalar field fluctuations are infinitely
growing in the course of time. In turn, this means that the
considered configuration is {\em unstable} against linear
spherically symmetric perturbations.

\section{Conclusions\label{VI}}
In this paper we have obtained exact non-static spherically
symmetric solutions in the theory of gravity with the scalar field
possessing the exponential potential. The solution \Ref{nonstatwh}
of particular interest corresponds to the scalar field with
negative kinetic energy, i.e. the ghost, and represents two
asymptotically homogeneous spatially flat universes connected by a
throat. In the other words, one may interpret such the spacetime
as a wormhole in cosmological setting. It is important to notice
that both the universes and the throat of the wormhole are
simultaneously expanding with acceleration. The character of
expansion qualitatively depends on the wormhole's mass parameter
$m$. In case $m=0$ the expansion goes exponentially, so that the
corresponding spacetime configuration, given by the metric
\Ref{metricm0}, represents two de Sitter universes joining by the
throat. In case $m>0$ the expansion has the power character, so
that the metric \Ref{powerlaw} describes now the inflating
wormhole connecting two homogeneous spatially flat universes
expanding according to the power law into the final singularity.

The stability analysis of the non-static wormholes has revealed
their instability against linear spherically symmetric
perturbations. This result is especially interesting in comparison
with the fact that static phantom wormholes are stable in this
case (see \cite{stability,Arm}). Thus, one may suppose that the
time dependence makes phantom wormholes to be unstable.

\section*{Acknowledgments}
S.S. acknowledge kind hospitality of Institute of Theoretical
Physics (Chinese Academy of Science). This project was in part
supported by National Basic Research Program of China under Grant
No. 2003CB716300 and by NNSFC under Grant No. 90403032. S.S. was
also supported in part by the Russian Foundation for Basic
Research grants No. 05-02-17344, 05-02-39023.

\section*{Appendix}
In the static spherically symmetric case $\phi=\phi(r)$ and the
spacetime metric can be taken as
\beq\label{generalmetric}
ds^2=-Adt^2+A^{-1}{dr^2}+Br^2(d\theta^2+\sin^2\theta d\phi^2),
\eeq
where $A$ and $B$ are two unknown functions of $r$. The system of
equations \Ref{sys} gets now the following form: \bea
\frac{A''}{A}+\frac{A'B'}{AB}+\frac{2A'}{rA}&=&0,\label{sys1a}\\
-\frac{A''}{2A}-\frac{B''}{B}-\frac{2B'}{rB}-
\frac{A'B'}{2AB}-\frac{A'}{rA}+\frac{B'^2}{2B^2}&=&\epsilon\phi'^2,\label{sys1b}\\
\frac{A'B'}{AB}+\frac{2A'}{rA}+\frac{B''}{B}+\frac{4B'}{rB}-\frac{2}{r^2AB}+\frac{2}{r^2}&=&0,\label{sys1c}
\eea
and
\beq\label{sysphi}
\epsilon\left[\phi''+\(\frac{A'}{A}+\frac{B'}{B}+
\frac2r\)\phi'\right]=0
\eeq
where a prime means the derivative
with respect to $r$. By integrating the equation \Ref{sys1a} we
find
\beq\label{relB}
B=\frac{b_0}{r^2A'},
\eeq
where $b_0$ is a constant of integration. Also, taking into
account Eq. \Ref{sys1a}, we can integrate the field equation
\Ref{sysphi}. As a result we obtain $\phi=\phi_1+\phi_0\ln A$,
where $\phi_0$ and $\phi_1$ are two constants of integration.
Without loss of generality we can put $\phi_1=0$, since the action
\Ref{action} is invariant with respect to the shift
$\phi\to\phi+\phi_1$. Now we have
\beq\label{relphi}
\phi=\phi_0\ln A.
\eeq
The relations
\Ref{relB} and \Ref{relphi} demonstrate to us that $\phi(r)$ and
$B(r)$ are expressed via $A(r)$. To obtain an equation for $A$ we
substitute Eqs. \Ref{relB}, \Ref{relphi} into \Ref{sys1b} and find
after some algebra
\beq\label{eqA}
\(\frac{A''}{A'}\)'-\frac12\(\frac{A''}{A'}\)^2=\lambda\(\frac{A'}{A}\)^2,
\eeq
where $\lambda\equiv \epsilon\phi_0^2$ is a new parameter with
values lying in the interval $(-\infty,\infty)$ depending on
$\epsilon$ and $\phi_0$. Note that $\phi_0=|\lambda|^{1/2}$, and
hence $\phi=|\lambda|^{1/2}\ln A$. A general solution of Eq.
\Ref{eqA} reads
\beq\label{gensolA}
A(r)=\exp\left\{\frac{2}{\sqrt{1+2\lambda}}\mathop{\rm
arctanh}\left[\frac{C_1(r+C_2)}{2\sqrt{1+2\lambda}}\right]+C_3\right\},
\eeq
where $C_1$, $C_2$, and $C_3$ are constants of integration. To
analyze the obtained solution it will be more convenient to
separate three cases depending on values of $\lambda$. Namely,

\vskip3pt{\bf Case I.} $\lambda>-1/2$. In this case the solution
\Ref{gensolA} reduces to
$$A(r)=A_1\(1-\frac{2m}{\delta(r-r_1)}\)^\delta,$$
where $\delta=(1+2\lambda)^{-1/2}$, and $A_1$, $m$, and $r_1$ are
arbitrary constants. By making an appropriate rescaling of
coordinates $t$ and $r$ we can put $A_1=1$ and $r_1=0$, so that
$A=(1-2m/\delta r)^\delta$. Now from Eq. \Ref{relB} we obtain
$B=(b_0/2m)(1-2m/\delta r)^{1-\delta}$. The value of $b_0$ is not
free. To fix it we should substitute $A(r)$ and $B(r)$ into Eq.
\Ref{sys1c}. This yields $b_0=2m$. Remembering that
$\phi=|\lambda|^{1/2}\ln A$ we come, finally, to the following
result: \bs\label{acaseI}
\bea &&ds^2=-\(1-\frac{2m}{\delta r}\)^\delta
dt^2+\(1-\frac{2m}{\delta r}\)^{-\delta}dr^2+\(1-\frac{2m}{\delta
r}\)^{1-\delta}r^2
(d\theta^2+\sin^2\theta d\varphi^2),\\
&&\phi(r)=\frac{|\lambda|^{1/2}}{(1+2\lambda)^{1/2}}\,\ln\(1-\frac{2m}{\delta
r}\),
\eea
\es
where $\delta=(1+2\lambda)^{-1/2}$. Note that the solution in the
form \Ref{acaseI} has been given by Buchdahl \cite{Buc}.

\vskip3pt{\bf Case II.} $\lambda=-1/2$. In this case the solution
\Ref{gensolA} reads
$$ A(r)=A_1e^{-2m/(r-r_1)},$$
where $A_1$, $m$, and $r_1$ are arbitrary constants. An
appropriate rescaling coordinates $t$ and $r$ yields
$A=e^{-2m/r}$. The functions $B(r)$ and $\phi(r)$ are found in the
same way as above. Finally, we obtain
\bs\label{acaseII}
\bea
&&ds^2=-e^{-2m/r}dt^2+e^{2m/r}[dr^2+r^2(d\theta^2+\sin^2\theta
d\varphi^2)],\\
&&\phi(r)=-\sqrt{2}\,\frac{m}{r}.
\eea
\es
First this solution has been exhibited by Yilmaz \cite{Yil}.

\vskip3pt{\bf Case III.} $\lambda<-1/2$. In this case the solution
\Ref{gensolA} takes the following form:
$$
A(r)=A_1\exp\left\{\frac{2}{|1+2\lambda|^{1/2}}\arctan\left(\frac{r-r_1}{r_0}\right)\right\},
$$
where $A_1$, $r_0$, and $r_1$ are arbitrary constants. Analogously
to the previous cases, appropriate rescaling coordinates $t$ and
$r$ yields $A_1=\exp(-m\pi/2 r_0)$ and $r_1=0$, so that
$A(r)=e^{2mu(r)}$, where we have denoted
$u(r)=(1/r_0)[\arctan(r/r_0)-\pi/2]$, and $
m/r_0=|1+2\lambda|^{-1/2}$. Now we find $B(r)$ and $\phi(r)$ and
come to the solution
\bs
\bea
&&ds^2=-e^{2mu(r)}dt^2+e^{-2mu(r)}[dr^2+(r^2+r_0^2)(d\theta^2+\sin^2\theta
d\varphi^2)],\\
&&\phi(r)=\frac{2|\lambda|^{1/2}}{|1+2\lambda|^{1/2}}\,\left(\arctan\frac{r}{r_0}
-\frac{\pi}{2}\right). \eea
\es

\end{document}